\documentclass{article}

\usepackage{arxiv}

\usepackage[utf8]{inputenc} 
\usepackage[T1]{fontenc}    
\usepackage{hyperref}       
\usepackage{url}            
\usepackage{booktabs}       
\usepackage{amsfonts}       
\usepackage{nicefrac}       
\usepackage{microtype}      
\usepackage{lipsum}
\usepackage{color}
\usepackage{algorithm}
\usepackage{algorithmic}
\usepackage{amsmath,amsfonts,amssymb}
\usepackage{subfigure}
\usepackage{lineno}
\usepackage{graphicx}
\usepackage{bm}
\usepackage{subfigure}

\title{Convolutional causal learning\\
for aerodynamic flows}

\author{
Ryo Koshikawa$^{[1]}$, Ryo Araki$^{[2]}$, Qiong Liu$^{[3]}$, and Kai Fukami$^{[1,*]}$
\\
\\
$[1]$ Department of Aerospace Engineering, Graduate School of Engineering, Tohoku University, Sendai, 980-8579, Japan\\
$[2]$ Department of Mechanical and Aerospace Engineering, Tokyo University of Science, Noda, 278-8510, Japan\\
$[3]$ Department of Mechanical and Aerospace Engineering, New Mexico State University, NM 88003, USA\\
$^{*}$Corresponding author: kfukami1@tohoku.ac.jp
}

\begin{document}
\maketitle

\begin{abstract}

\vspace{-2mm}
{This study {aims to capture} aerodynamic causality from snapshot data with a time-varying mode decomposition technique referred to as information-theoretic machine learning.
The current approach extracts time-dependent informative vortical structures, contributing to the future evolution of the aerodynamic coefficients. 
The present decomposition is employed with a convolutional neural network, enabling the identification of the spatial continuous mode. 
In addition, a low-order representation, characterizing the informative vortical structures and their corresponding aerodynamic coefficients, can also be identified by considering autoencoder-based data compression.
The present technique is applied to a range of aerodynamic examples, including extreme vortex-gust airfoil interactions, experimentally measured transverse jet-wing interaction, and a turbulent separated wake {across different Reynolds numbers.}
For the cases of gust-wing interaction, the time-varying gust effect on the lift response is extracted in an interpretable manner. 
With the example of a turbulent wake, the relationship between large-scale vortical motion and lift force is identified without any spatial length-scale information.  
The proposed approach could serve as a foundation for data-driven causal modeling and control for a range of unsteady flows. }

\end{abstract}

\section{Introduction}
\label{sec:intro}
\vspace{-2mm}

{
Understanding the interaction between vortical structures and the exerted force response is essential for efficient flow control and modeling of aerodynamic flows.
The process by which the surrounding flows induce forces is regarded as a causal relationship, where the vortical structures act as the cause and the aerodynamic force as the effect.
This study discusses how such an aerodynamic relationship can be captured in a data-driven manner.

To examine the contribution of specific vortical structures to aerodynamic forces, a range of approaches have been considered, including the force elemental method~\cite{chang1992potential} and the vortex force map approach~\cite{otomo2025vortex}, which provide spatial locations of force-generating structures. 
Modal analyses, such as proper orthogonal decomposition ~\cite{lumley1967structure} and dynamic mode decomposition~\cite{schmid2010dynamic}, have been widely employed to extract dominant coherent structures from fluid flows. 
Such techniques can serve to compress the inherently high-dimensional flow data as a foundation for reduced-order modeling, which is followed by nonlinear machine learning techniques, capturing intrinsic nonlinearity into a low-dimensional representation~\cite{brunton2020machine,taira2025machine}.

{
 Data-driven analysis for fluid dynamics has evolved to address complex flow physics in several directions. For example, to extract coherent structures that are highly correlated with a specific target variable, such as aerodynamic forces, extended proper orthogonal decomposition has been {considered}~\cite{boree2003extended,discetti2019characterization}. 
To capture the nonlinear temporal evolution of such dynamics, time-delay embedding~\cite{takens2006detecting,arbabi2017ergodic,bakarji2023discovering} can be employed, elevating modal analysis from feature extraction to the prediction of complex dynamics~\cite{schmid2022dynamic}.
Furthermore, spectral linear stochastic estimation ~\cite{adrian1988stochastic,tinney2006spectral} provides the mean square linear estimate of a complex turbulent component from a predictor field based on the correlation, which has been applied to identify large-scale coherent structures in turbulent jets~\cite{tinney2006spectral}.
While conventional data-driven modal analyses are primarily designed for statistically steady flows, recent studies have presented that data-driven time-dependent bases analysis enables the analysis of transient flow dynamics with an unsteady base state~\cite{ashtiani2022scalable,ashtiani2025data}, which has also been examined with operator-based techniques~\cite{babaee2016minimization,kern2024onset,zhong2025optimally}.
}

Moving beyond correlation-based analysis, a causality-inspired approach has recently emerged to identify the causal drivers of unsteady flows, thereby assisting in a deeper understanding of complex fluid motion and dynamics.
For example, the use of the Shapley additive explanations (SHAP) algorithm for turbulent flow prediction has been shown as an effective data-driven approach to identify vortical structures~\cite{cremades2025classically, cremades2026x}, {with respect to} {the conveyed information about the temporal flow development.}
Causality-based modal analysis, referred to as informative and non-informative mode decomposition, has also been introduced~\cite{arranz2024informative}. {The formulation has been extended to aerodynamic flows by considering the lift-generating mechanism as a cause and effect relationship ~\cite{fukami2025information}}. It is important to note that such approaches enable the modal analysis of flow with aperiodic or transient base states, while traditional techniques are often limited to the dynamics with a statistically-stationary base flow~\cite{linot2025extracting}.
{
Furthermore, the inherent nonlinearity offers an advantage to capture dominant features in a nonlinear system over the conventional linear method~\cite{arranz2024informative}.}

{
This study considers {\it information-theoretic convolutional learning} that achieves mode decomposition based on the informatics point of view, providing modal structures related to aerodynamic force while identifying reduced-order representations of important vortical structures.}
While existing information-theoretic mode decompositions operate in a point-wise manner, which yields spatial discontinuity along with an expensive inference cost~\cite{fukami2025information}, the current technique, based on convolutional networks~\cite{lecun2002gradient}, enables the extraction of coherent modal structures and identifies a submanifold that represents the relationship between vortical flows and aerodynamic forces.
The proposed method is applied to a range of aerodynamic flow examples, exhibiting that the present information-theory-assisted model is capable of extracting the relationship between vortical structures and exerted lift in a transient manner.

The present paper is organized as follows.
The formulation of informative mode decomposition is described in section \ref{sec:method}. 
Results are discussed in section \ref{sec:res}. Conclusions are remarked in section \ref{sec:conc}.
\section{Approach}
\label{sec:method}
\begin{figure}
    \begin{center}
        \includegraphics[width=1\textwidth]{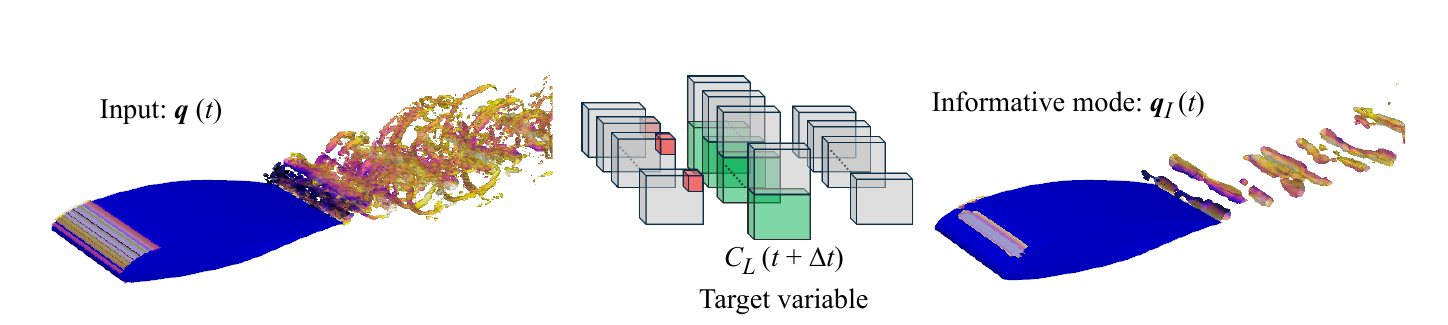}
    \end{center}
    
    \vspace{-2mm}
    \caption{
    An example of the given state $\bm q$ and the informative component $\bm q _I$ decomposed by a data-driven technique.
    }
    \vspace{-3mm}
    \label{fig1}
\end{figure}

This study aims to capture the causal relationship between vortical flows and aerodynamic response in a data-driven manner.}
To achieve this, we consider decomposing a given state ${\bm q}({\bm x},t)$ based on the contribution to the target variables at a future time step $\bm{\lambda}(\bm{x},t+\Delta t)$. 
The present approach, illustrated in figure~\ref{fig1}, decomposes a given flow snapshot in a manner,
\begin{equation}
    \bm{q}(\bm x, t) = \bm{q}_I(\bm x, t) +\bm{q}_R(\bm x, t), 
\end{equation}
where ${\bm{q}}_{I}$ and ${\bm{q}}_{R}$ are the informative and residual components, respectively~\cite{arranz2024informative}.
In this study, the field of vorticity $\bm \omega$ and the second invariant $Q$ of the velocity gradient tensor are considered as the given state $\bm q(\bm x, t)$, while the lift coefficient $C_L \equiv F_L/(0.5\rho u_\infty ^2 c)$ is selected as the target variable $\bm \lambda$.
Here, $F_L$, $u_\infty$, $c$, and $\rho$ denote the lift force, freestream velocity, chord length, and fluid density, respectively.
This setup is motivated by the fundamental aerodynamic relationship between lift and circulation $\Gamma$, i.e., $\Gamma \propto C_L$, which may enable us to examine how the present method captures such underlying physics in a data-driven manner.

While traditional data-driven approaches extract dominant features based on correlation, {this study aims to perform modal extraction based on the pre-defined causal relationship.}
{To achieve this, we consider measuring the amount of information quantitatively with the concept of Shannon entropy, which assesses uncertainty and randomness of arbitrary variables~\cite{shannon1948mathematical,martinez2024decomposing}, following the formulation of informative and non-informative decomposition originally introduced by Arranz \& Lozano-Duran~\cite{arranz2024informative}.
The Shannon entropy $H(\bm{\lambda})$ for a target variable $\bm {\lambda}$ in the future is described as
\begin{align}
    H(\bm{\lambda})=-\sum_{\bm S \in \mathcal{S} } p_{\bm{\lambda}}(\bm{\lambda}=\bm S)\log p_{\bm{\lambda}}(\bm{\lambda}=\bm S),
\end{align}
where $p_{\bm{\lambda}}$ is the probability of $\bm \lambda$ being in state $\bm S$ and $\mathcal{S}$ represents the set of all possible states of $\bm \lambda$. 
The remaining information about $\bm {\lambda}$, not contained in $\bm q_I$, is measured by the conditional Shannon entropy described by
\begin{align}
    H(\bm{\lambda}|\bm{q}_I)=-\sum_{\bm S \in \mathcal S} \sum_{\bm R \in \mathcal R} p_{\bm{\lambda},\bm{q}_I}(\bm S, \bm R)\log \frac{{p_{\bm{\lambda}, \bm q_I}(\bm S, \bm R)}}{p_{\bm{q}_I}(\bm R)},
\end{align}
where $p_{\bm \lambda, \bm q_I}$ is the joint probability distribution of $\bm \lambda$ and $\bm p$, $\bm R$ is a state of the informative component $\bm q_I$, and $\mathcal R$ is all the possible states of $\bm q_I$. 
The difference between $H(\bm{\lambda}|\bm{q}_I)$ and $H(\bm{\lambda})$ is called the mutual information such that $ I(\bm \lambda; \bm q) = H(\bm{\lambda}) - H(\bm{\lambda}|\bm{q}_I)$, expressing the amount of information shared between $\bm q_I$ and $\bm \lambda$.
The informative component is defined as the state that maximizes mutual information with $\bm \lambda$ at a future time stamp. 
This is achieved when the conditional Shannon entropy of $\bm \lambda(t+\Delta t)$ and the informative component $\bm q_I$ is zero, $H(\bm{\lambda}|\bm{q}_I) = 0$,
which means $\bm q_I$ completely determines the target variables $\bm{\lambda}$.
Furthermore, the mutual information between the informative and residual components should be zero, 
\begin{equation}
    I(\bm q_R; \bm q_I) = 0,
\end{equation}
since each decomposed component should be statistically independent.

\begin{figure}
    \centering
    \includegraphics[width=1\textwidth]{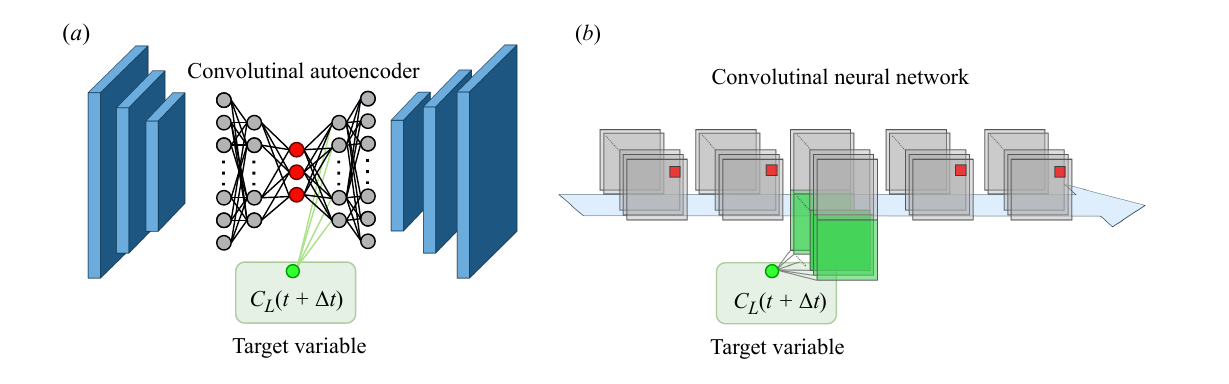}
    \vspace{-6mm}
    \caption{
    Informative mode extractor $\mathcal F$ based on ($a$) convolutional autoencoder and ($b$)~convolutional neural network.}
    \vspace{-3mm}
    \label{fig2}
\end{figure}
To extract the informative vortical structure $\bm q_I$ from the given vorticity field $\bm q$ with respect to a future target of lift, i.e., $\lambda = C_L$, we construct an informative mode extractor $\mathcal F$,
\begin{align}
    \bm q_I(t) = \mathcal{F}(C_L(t+\Delta t), \bm q(t); \bm w),
\end{align} 
where $\bm w$ is the weight parameter of the mode extractor. 
This extractor is implemented as a certain type of neural network, convolutional deep sigmoidal flow \cite{huang2018neural}.
This model is constrained to possess non-negative weight values with bijective activation functions, which guarantees a bijective transformation to offer the solution $\bm q _I$ satisfying an information-theoretic condition of $H(C_L|\bm{q}_I) = 0$.
This extractor decomposes the given state at each time stamp, yielding time-varying informative modes.
The optimization for the weight distribution is performed with 
\begin{align}
    \bm w^* = \operatorname{argmin}_{\bm w}||\bm q - \bm q_I||_2+\beta||I(\bm q_R; \bm q_I)||_2,
    \label{eq:costfunc}
\end{align}
where the regular regression loss and mutual information loss are balanced with a constant parameter $\beta$~\cite{arranz2024informative}.
{Without the second term, the present model simply replicates the input field, i.e., $\bm q_I \approx \bm q$.}
{The value of $\beta$ is determined based on the L-curve analysis~\cite{hansen1993use}, facilitating the identification of the trade-off relationship between two terms, which will be discussed later.}
By minimizing the above cost function, the model is designed to tune $\bm w$ to extract vortical structures $\bm q_I$ that contain information about aerodynamic response.
{Note that the current formulation differs from conventional definitions of causal inference~\cite{imbens2015causal} as this study does not consider any manipulations or perturbations in the system.
Rather, this study focuses on the force generation dynamics due to the presence of vortical structures.
}

This study chooses two neural-network architectures as a mode extractor, depending on the flow of interest.
We first consider a convolutional autoencoder-based model~\cite{lecun2002gradient, hinton2006reducing}, as illustrated in figure~\ref{fig2}($a$).
This is used for cases where the embedding latent dimensions are recognized to be few, i.e., ${\mathcal O}(10^0)$.
In other words, the model provides a low-order representation, in addition to performing the current decomposition.
A series of convolutional neural networks without compression is then employed for the analysis of turbulent vortical structures, as shown in figure~\ref{fig2}($b$).

Distinct from previous studies performing the information-based decomposition via multi-layer perceptrons that require flattening of the input data~\cite{ arranz2024informative,fukami2025information}, both models in this study are based on the convolutional operation, allowing the decomposition across the entire snapshot with one shot while preserving the spatial arrangement of vortical structures~\cite{morimoto2021convolutional}.
This mitigates an issue of point-wise decomposition, often yielding spatially-discontinuous modal structures~\cite{fukami2025information}.
{These information-theoretic models are trained to learn the relationship between two inputs: the current flow field snapshot ${\bm q}(t)$ and the lift coefficient over the time interval $\Delta t$, ${C_L}(t+\Delta t)$. In other words, the models are designed to remove redundant structures, which do not contribute to the lift after the time gap $\Delta t$, thereby providing the time-varying informative mode ${\bm q}_I(t)$, expected to vary with the value for~$\Delta t$.
We also discuss the dependence of informative mode on~$\Delta t$.}
 
{Note that ``mode" produced with the current technique differs from the time-independent basis in linear modal analyses. 
The current informative mode ${\bm q}_I(t)$ would be analogous to modes multiplied or amplified with time-varying temporal coefficients in a linear context.
While some previous studies have examined the isolation of explicit nonlinear bases from the internal architecture of machine-learning models~\cite{erichson2020shallow}, we hereafter follow previous studies that refer to nonlinearly reconstructed flow fields, obtained from the time-varying coefficients and the basis of a nonlinear network, as ``modes"~\cite{murata2020nonlinear,eivazi2022towards,cremades2026x,fukami2020convolutional}.}
\section{Results}
\begin{figure}
    \centering
    \includegraphics[width=1\textwidth]{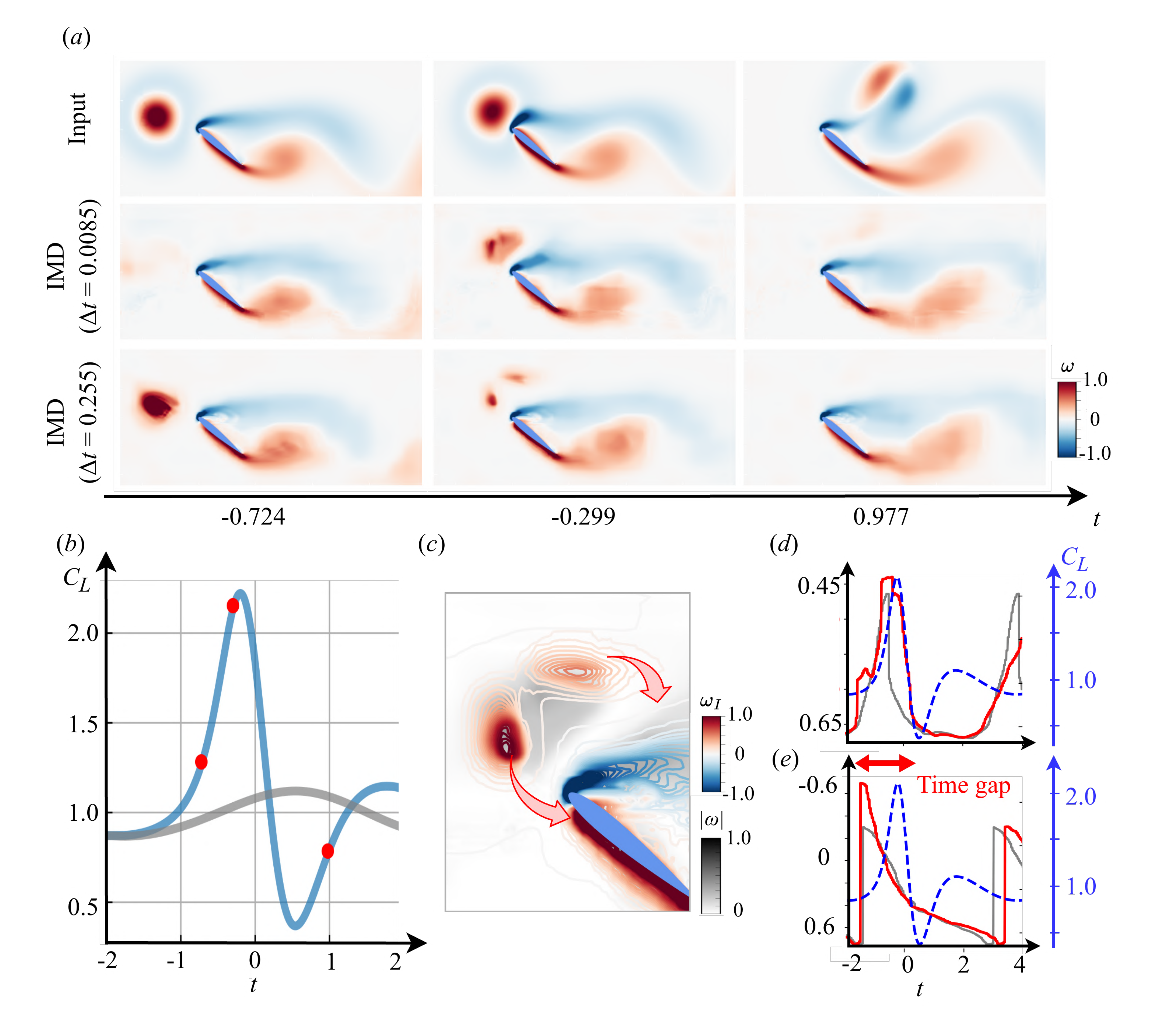}
    \vspace{-7mm}
    \caption{
    Informative mode decomposition of extreme vortex-airfoil interaction. $(a)$~Vorticity field (Input) and extracted modes (IMD). $(b)$~Time trace of lift coefficient, where convective time is set to be zero when the gust center reaches the leading edge (gray line: undisturbed case). $(c)$ The zoomed-in view of extracted mode at $t = -0.299$ with $\Delta t=0.255$. Latent-variable evolution with ($d$) $\Delta t = 0.0085$ and $(e)$~$\Delta t = 0.255$. 
    }
    \vspace{-7mm}
    \label{fig3}
\end{figure}
\label{sec:res}
{
To examine how the current informative mode decomposition (IMD) extracts time-varying modes based on the {contribution to the future lift response}, we consider three examples of aerodynamic flows, {covering a range of spatiotemporal complexity across the chord-based Reynolds number~$Re$:}
($\mathrm{i}$) extreme vortex gust-airfoil interactions at $Re~=~100$ \cite{fukami2023grasping},
($\mathrm{ii}$) experimentally-measured transverse gust-wing interactions at $Re = 20,000$ \cite{towne2023database},
and ($\mathrm{iii}$) a separated turbulent wake over a wing section at $Re = 20,000$ \cite{liu2025design}.

The first example with a discrete vortex gust highlights the applicability of the present method in capturing highly unsteady transient dynamics.
With the second example of experimental datasets, the robustness against experimental noise of the present model is examined.
We also consider a separated turbulent wake in a three-dimensional domain under quasi-cyclic behavior, discussing how the underlying physics is extracted for aerodynamics scenarios exhibiting a range of spatial length scales.

\subsection{Example 1: extreme vortex-gust airfoil interactions}
Let us consider an extreme vortex gust-airfoil interaction around an NACA0012 airfoil with an angle of attack $\alpha = 40^\circ$ at $Re =100$.
Data sets are produced by fully-validated and verified direct numerical simulations~\cite{fukami2023grasping,fukami2024data}. 
Unsteady periodic shedding is observed for the case without the presence of a vortex gust.
A Taylor vortex~\cite{taylor1918dissipation} is introduced upstream of an airfoil, producing transient and drastic excitation of aerodynamic characteristics.
Gust ratio $G\equiv u_{\theta,\rm max}/u_{\infty}$, where $u_{\theta,\rm max}$ is the maximum rotational velocity, and the gust diameter $D$ normalized by the chord length are set to $(G,D) = (2,0.5)$.
Note that the current condition with $|G|>1$, which can happen in wind shear, mountain-covering areas, and severe atmospheric turbulence, is classified as {\it extreme aerodynamics}, traditionally thought of as unflyable for small-sized aircraft~\cite{jones2022physics,taira2026extreme,fukami2023grasping}.
Understanding the process of lift generation in such transient dynamics is crucial for achieving stable flight operations. 
This example, hence, can serve as an ideal case to examine how the model identifies time-varying relationships between vortical motion and aerodynamic response.

We perform the present decomposition on the vorticity field $\bm{\omega} (\bm x,t)$, as exhibited in figure~\ref{fig3}.
Here, the balancing parameter $\beta$ is set to $0.05$, {although the effect of $\beta$ will be examined later}.
The present convolutional learning-based approach offers time-varying modes that capture the effect of gust on the lift response, as shown in figure~\ref{fig3}($a$).
With $\Delta t = 0.0085$, where an almost instantaneous {contribution} is considered, the current approach offers a time-dependent informative mode capturing the influence of the gust on the lift response, as shown in figure~\ref{fig3}($a$).
At $t = -0.724$, when the gust is still distant from the leading edge, the gust does not appear in the informative mode, implying that the temporal lift variation is determined by flow around a wing, rather than the gust itself.
Strong impingement of the counter-clockwise vortex near the leading edge introduces a sharp peak in the lift response, as presented in figure~\ref{fig3}($b$).
Near the peak at $t = -0.299$, the gust appears in the extracted mode since the effect of the gust becomes dominant.
In contrast, when the lift fluctuation heads to the level of the undisturbed scenario at $t = 0.977$, the separating structure is recognized as less informative than the structure around the airfoil.
The presence of the gust in the informative mode is observed within a limited time duration, exhibiting that the exerted lift is governed by a finite time {interaction}.

The dependence of the informative mode on the time interval $\Delta t$ is then examined.
With a larger time window of $\Delta t = 0.255$, the vortex core appears in informative mode at $t = -0.724$, whereas the model with $\Delta t = 0.0085$ disregards it.
Furthermore, the vortex core appears to split into left and right halves near the leading edge at $t = -0.299$, as highlighted in figure~\ref{fig3}($c$).
The left part is also highlighted in the informative mode with $\Delta t = 0.0085$, suggesting that this structure directly affects the airfoil.
On the other hand, the right part, which disappears $0.7$ convective time later than the left one, seems to contribute to the lift response by interacting with the separated wake.
This suggests that the model captures two types of lift generation processes in the current transient aerodynamic scenario.

\begin{figure}
    \centering
    \includegraphics[width=1\textwidth]{./Fig4.pdf}
    \vspace{-7mm}
    \caption{
    {
    The dependence of the mutual information loss on the number of training snapshots $n_{\rm snapshot}$ among all the snapshots $n_{\rm all}$. Shown on the right are informative components at $t = -0.299$.
    }}
    \vspace{-4mm}
    \label{fig_snap}
\end{figure}
\begin{figure}
    \centering
    \includegraphics[width=1\textwidth]{./Fig5.pdf}
    \vspace{-5mm}
    \caption{
    {
    The dependence of the informative components on the value of $\beta$. The decomposed informative modes at $t = -0.299$ are shown with the L-curve plot.
    }
    }
    \vspace{-3mm}
    \label{fig:beta}
\end{figure}

A low-dimensional representation identified through the present autoencoder-based model is further analyzed in figures~\ref{fig3}($d$) and ($e$).
The latent dimension is set to be three, following the previous study that discusses the manifold discovery of extreme vortex-airfoil interactions~\cite{fukami2023grasping}.
The red trajectory for the disturbed case deviates from the gray one for the undisturbed flow and shows peaks, highlighting when the effect of the gust is prominent.
With $\Delta t = 0.0085$, these peaks coincide with those of the lift coefficient induced by interaction with the gust, suggesting that the current latent space contains an instantaneous {relationship with aerodynamic response}. On the other hand, with $\Delta t = 0.225$, time gaps are observed between these peaks.
The assessment of the gust contribution to the lift response is embedded in the low-order representation based on the {information about the target variable}.
The other two latent variables exhibit a similar trend, although not shown here.
The proposed method learns {relationship between vortical motions and future lift response} in a transient manner and extracts coherent structures as a time-varying mode in highly unsteady aerodynamic environments.

{
To extract informative components from snapshot data by learning the relationship between vortical structures and future lift response, {an appropriate amount of training snapshots needs to be }provided to the present model. The influence of the number of snapshots on the mutual information loss $I(\bm q_R; \bm q_I)$ is evaluated for $n_{\rm snapshot} /n_{\rm all} = 1/16$ to $1$ in figure~\ref{fig_snap}, where $n_{\rm all} (=2400)$ represents the total number of prepared snapshots. 
In general, the mutual information loss increases as $n_{\rm snapshot}$ decreases, indicating that {the resulting extraction} is affected by the decrease in the training data set. Here, we also examine the decomposed informative components for $n_{\rm snapshot} /n_{\rm all} = 1/16$ and $1/8$ with different values for the balancing parameter $\beta$. 
While the extracted mode with $n_{\rm snapshot}/n_{\rm all} = 1/8$ and $\beta = 0.001$ exhibits high mutual information loss and deformed structures, by carefully tuning $\beta$, the present model is capable of extracting vortical structures with a mutual information loss comparable to that obtained using all snapshots, thereby identifying similar structures.
}

\begin{figure}
    \centering
    \includegraphics[width=1\textwidth]{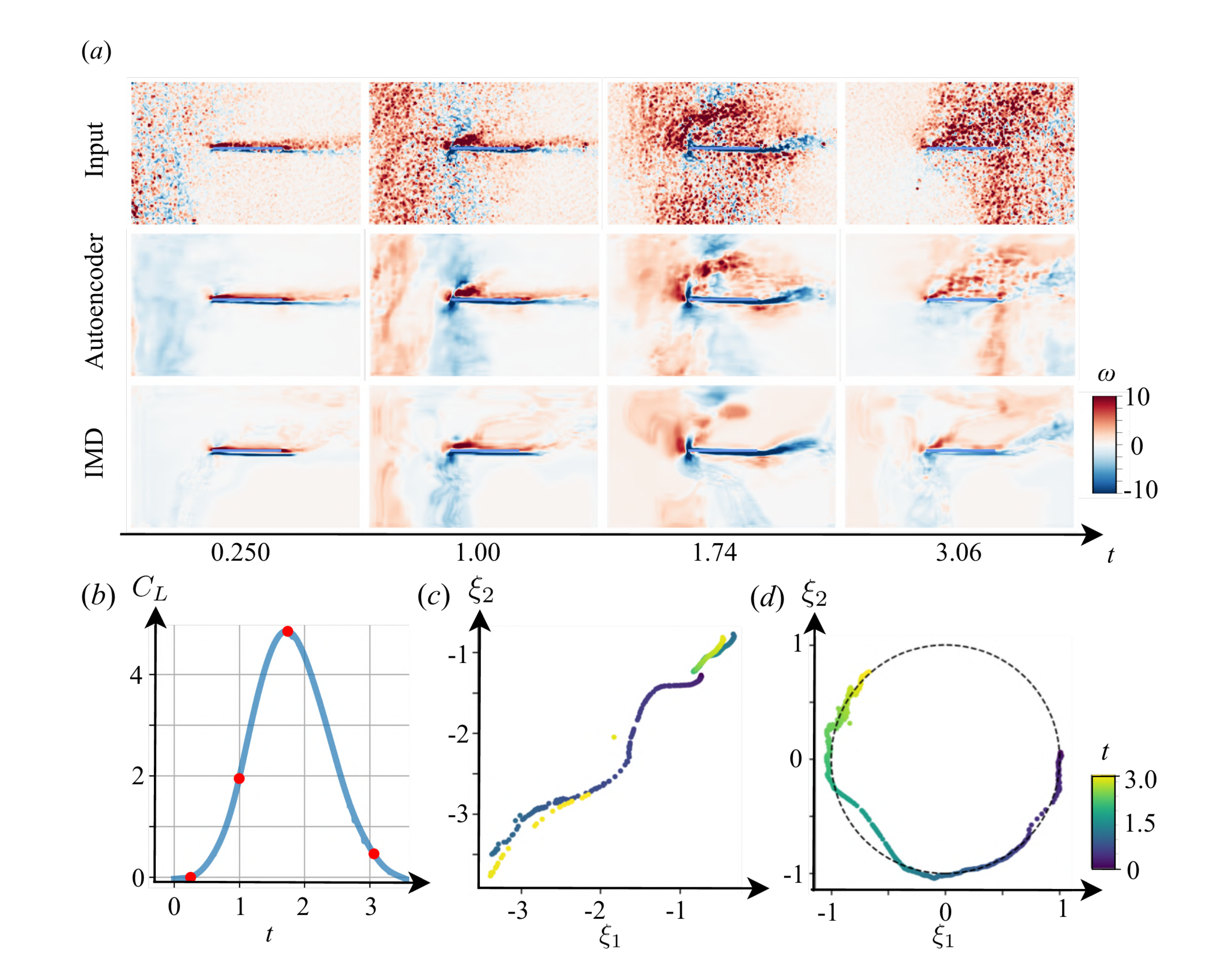}
    \vspace{-6mm}
    \caption{
    Informative mode decomposition for experimental transverse gust encounter at $Re = 20, 000$. ($a$) Vorticity snapshots, reconstructed flow field via convolutional autoencoder, and extracted informative fields. ($b$) Time series of lift coefficient. Latent space identified by the models ($c$) without and
($d$) with additional geometric constraints. 
    }
    \vspace{-2mm}
    \label{fig4}
\end{figure}

{
To determine the value of the balancing parameter $\beta$, we employ L-curve analysis~\cite{hansen1993use}, facilitating the identification of the trade-off relationship between the two terms in equation~\ref{eq:costfunc}. The relationship between the reconstruction loss and the mutual information loss across several values of $\beta$, along with the resulting decomposed modes, is examined, as shown in figure~\ref{fig:beta}. With a small $\beta$ of $0.01$, the model prioritizes the regular regression loss, yielding modes almost identical to the given state, while increasing $\beta$ allows the model to output modal structures distinct from the input field. Excessive penalty for the mutual information term hinders the reconstruction. By following this procedure, the balancing parameter $\beta$ is systematically determined for each case.
}

\subsection{Example 2: experimental measurements of large-amplitude transverse gust encounter}

To examine the applicability of the current techniques to experimental measurements, the flow around a flat plate at a constant angle of attack $\alpha = 0^{\rm \circ}$ with $Re = 20,000$ passing through a large-amplitude transverse jet at gust ratio $G = 1.5$, made available by Towne et al.~\cite{towne2023database}, is considered.
Details on the data curation and experimental setups are referred to Biler et al.~\cite{biler2021experimental} and Andreu-Angulo et al.~\cite{andreu2020effect}.
Here, we consider the spanwise vorticity field $\bm{\omega} _z (\bm x,t)$ as a given state.
For this case, we also employ a convolutional neural network-based autoencoder without the {information-theoretic term}~\cite{fukagata2025compressing}.
This comparison is intended to distinguish the effects of denoising and {removal of redundant components}, as an autoencoder composed solely of regression loss can remove noise and extract coherent structures through the compression process, as shown in figure~\ref{fig4}($a$).
The difference in reconstruction between the regular autoencoder and the current method derives from introducing a mutual information term and target variable input, exhibiting how the inferred {mutual information} works in the present data-driven modal analysis.

Let us perform the present informative mode decomposition for the experimental data set.
Here, the parameters of the time window $\Delta t$ and the balancing parameter $\beta$ are set to $0.005$ and $1$, respectively.
The current model extracts coherent structures capturing the effect of aerodynamic response, as shown in figure~\ref{fig4}($a$).
At $t = 0.250$, when the lift coefficient is nearly zero, the model disregards the majority of the given state, including the approaching gust structures.
In other words, the lift force is recognized as statistically independent from the surrounding vortical flow.
When the large-scale separation occurs along with the formation of positive vorticity at the leading edge, the time series of the lift coefficient shows a sharp increase for $0.5<t<1.74$, as exhibited in figure~\ref{fig4}($b$). 
At $t = 1.74$, large-scale structures with positive vorticity at the leading edge and separated wake with negative vorticity at the trailing edge are assessed as informative.
Once the transverse gust transitions to turbulence and dissipates at $t = 3.06$, the fine-scale structures that appear above the flat plate are disregarded.
Similar to the case of the vortex-airfoil interaction, vortical structures modified extensively by interaction with the gust are dominant contributors to the lift response.

\begin{figure}
    \centering
    \includegraphics[width=1\textwidth]{./Fig7.pdf}
    \vspace{-4mm}
    \caption{
    {
    Probability density functions of the vorticity (gray: input, red: IMD) are shown with informative components at representative time $t = 0.250$, $1.00$, and $1.74$. 
    }
    }
    \vspace{-2mm}
    \label{fig_add_PIVPDF}
\end{figure}

Low-order representations are also examined. 
The latent dimension is set to be 2, following Smith et al.~\cite{smith2024cyclic}, showing that the transverse gust encounter is possible to be considered as a cyclic event and compressed into a circle-shaped latent space.
The compression result with the cost function in equation~\ref{eq:costfunc} is shown in figure~\ref{fig4}($c$).
The trajectory collapses into a line with temporal discontinuity, which is challenging to interpret.
To address this, we constrain the latent space to lie on a circle by adding a geometry-related cost function~\cite{smith2024cyclic}.
The resulting latent variables form a circular-like trajectory that evolves continuously over time, as shown in figure~\ref{fig4}($d$).
Note that despite such an additional constraint, the present model successfully extracts the informative structure in a similar manner to that shown above.
This indicates that the present approach allows identification of physically interpretable low-order representations in addition to providing time-varying modal structures based on the {relationship with the target variable}.

{
Let us examine the probability density function of the input vorticity and the decomposed field, p.d.f.$(w_z)$, as shown in figure~\ref{fig_add_PIVPDF}. 
The {probability density function} of informative components exhibits a sharper peak near zero compared to the input data, suggesting that experimental noise contained in the input snapshots is removed through the extraction process.
In addition, the tail of the probability density function for the informative component, corresponding to the extreme positive and negative vorticity, is shrunk compared to that for the input.
This implies that intensely rotating structures do not necessarily act as the dominant drivers of future lift.
Furthermore, at $t = 1.74$, when strong impingement causes a sharp excitation of the lift force, we observe that the informative components retain the negative values in the {probability density function}.
This indicates that the extraction process reconstructs the transverse jet reaching the lower surface of the flat plate, a lift-related phenomenon obscured by experimental noise. 
The current assessment highlights the capability of the present model to selectively preserve lift-related vortical structures while mitigating experimental noise.
}

\subsection{Example 3: separated turbulent wake over a wing section of NLF(1)-0115 airfoil}

\begin{figure}
    \centering
    \includegraphics[width=1\textwidth]{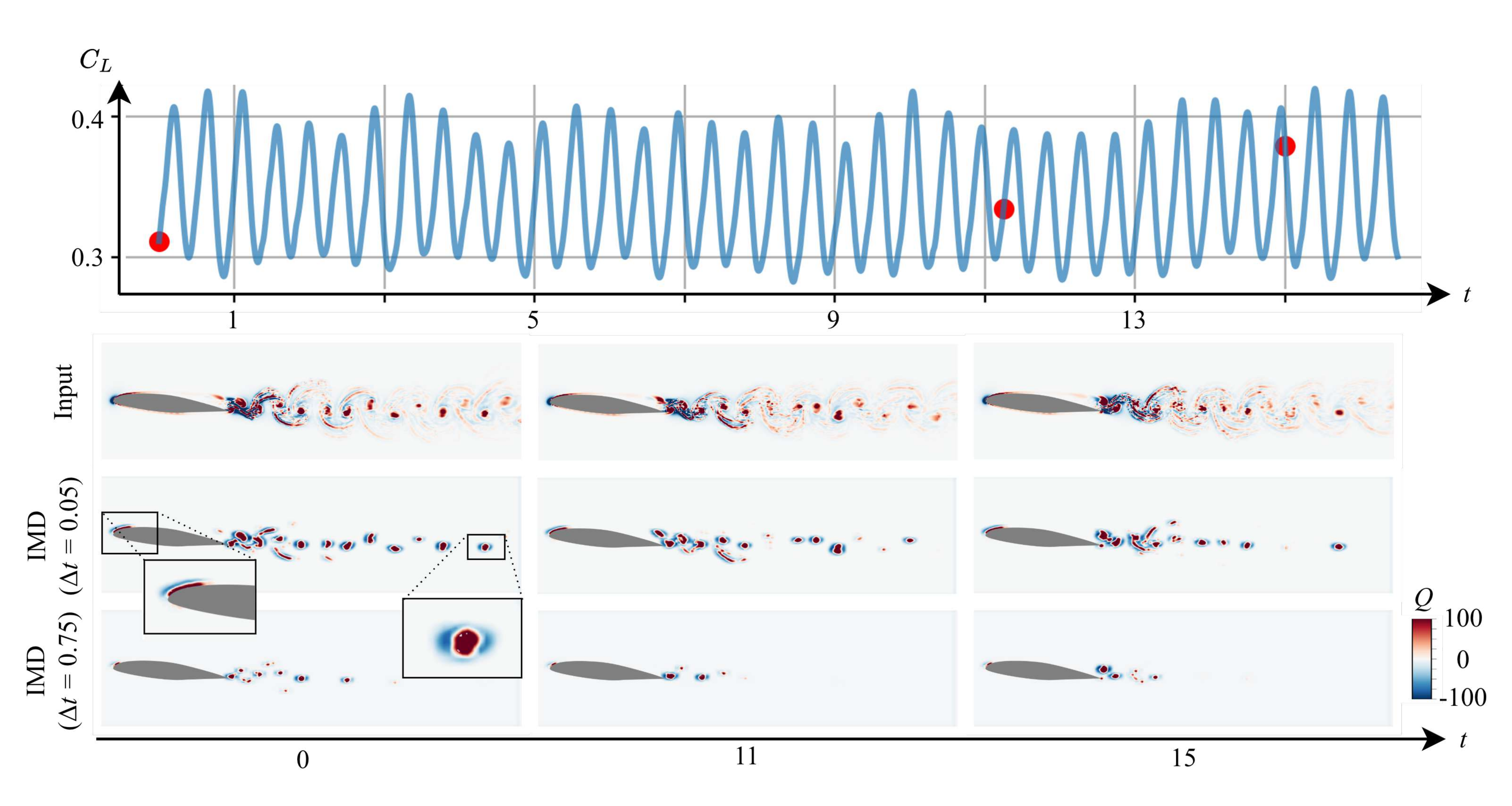}
    \vspace{-4mm}
    \caption{Informative modal structure of spanwise-averaged separated flow over wing section at $Re = 20, 000$ and dependence of decomposed mode on time window.
    }
    \vspace{-3mm}
    \label{fig5}
\end{figure}
Let us also examine turbulent flow with the current decomposition. Turbulent wake over a wing section, NLF(1)-0115 airfoil at an angle of attack of $\alpha = 5^\circ$ with $Re= 20,000$ is considered.
Separation occurs at the leading edge, producing a quasi-cyclic turbulent wake. 
We discuss how the proposed method captures informative structures responsible for lift with the presence of a range of scales in length. 
The datasets are produced by three-dimensional large-eddy simulations.
Further details on data curation are seen in Liu et al.~\cite{liu2025design}.
The field of the second invariant for the velocity gradient tensor, $\bm{Q}(\bm x, t)$, is considered as the source state.
In this example, a series of convolutional neural networks without data compression is employed, unlike the other two cases.
This is based on the prior knowledge that fine-scale structures could be lost due to two reasons: their small contribution to aerodynamic force~\cite{fukami2025information}, and the compression process inside the network~\cite{fukami2025extreme}.
By removing the compression layers inside the network, we aim to isolate the physical contribution of these structures.

The current decomposition is performed into the two-dimensional spanwise-averaged field based on an instantaneous and a delayed {contribution} where convective time windows $\Delta t$ are set to $0.05$ and $0.75$, respectively, as shown in figure~\ref{fig5}. The balancing parameter $\beta$ is set to $1 \times 10^4$.
For the case of $\Delta t = 0.05$, large-scale vortex cores and structures around the leading edge are primarily identified due to their significant contribution.
The result with $\Delta t = 0.75$ shows that the present model selectively isolates the contribution of the leading edge and the vortex core near the trailing edge to the lift response, while identifying the convection process of the vortex cores and shear to disregard them. 
Furthermore, the vortex cores in the informative mode are observed to be located between structures with negative $Q$.
This shear is captured mainly on both the upstream and downstream sides, indicating that the shear induced by the convection of the vortex core in the streamwise direction is recognized as informative in addition to the core itself.

The present decomposition is finally extended to the three-dimensional flow field, as shown in figure~\ref{fig6}($a$).
Similar to the spanwise-averaged case, the large-scale vortex cores are extracted as informative, while fine-scale structures are disregarded due to their small contribution to the lift.

\begin{figure}
    \hspace{-10mm}
    \includegraphics[width=1.1\textwidth]{./Fig9.pdf}
    \vspace{-4mm}
    \hspace{-5mm}
    \caption{
    ($a$) Informative modal structure of three-dimensional separated flow over wing section at $Re = 20, 000$ visualized with the iso-surface ($Q_{\rm th} = 100$) colored by streamwise velocity $u$. {($b$) Probability density function of $Q$-criterion field, p.d.f.($\bm Q$).} ($c$)~Scale-decomposed fields with two cuts of length scales.}
    \vspace{-3mm}
    \label{fig6}
\end{figure}

{
We next examine the probability density function of the reference and the decomposed $Q$-criterion field, p.d.f.$(\bm Q)$, as shown in figure~\ref{fig6}($b$). The probability function of the decomposed informative mode exhibits that the present model predominantly retains the structures with $Q > 0$. The current rotation-dominated extraction suggests that the present model assesses rotating structures as the primary drivers of aerodynamic lift generation.
It is also observed that the probability density of the informative component is lower than the reference at the extreme positive region, indicating that not all intensely rotating structures are regarded as informative for the future lift. The model selectively filters out these intense, yet with fewer contributing structures, and isolates lift-related structures.
}

The length scale of structures extracted through the current method for the three-dimensional case is further discussed through the scale-decomposition analysis~\cite{goto2017hierarchy,fujino2023hierarchy}. 
A spatial band-pass filter based on a two-dimensional Gaussian kernel $G$ is applied to the velocity field in the $x$ and $y$ directions. 
The cut-off length scale $\sigma_{\rm max}$ is set to $\sigma_{\rm max}=c/(2\pi St)$, which is the diameter of the roller vortices, shed with the Strouhal number $St$.
The operation is described as 
\begin{align}
\boldsymbol{u}^{[\sigma_1, \sigma_2]}(x, y, z, t) &= \int_{\mathcal D} \boldsymbol{u}(x, y, z, t) [G(x', y'; x, y, \sigma_1) - G(x', y'; x, y, \sigma_2)] dx' dy'
\nonumber
\end{align}
where $\mathcal D$ is the domain of integration.
The $Q$-criterion fields for a length scale range of [$\sigma_1, 2\sigma_1$] with $\sigma_1 =(\sigma_{\rm max}, \sigma_{\rm max}/6)$ are computed from scale-decomposed velocity fields, as exhibited in figure~\ref{fig6}($c$).

With $\sigma_1=\sigma_{\rm max}$, the scale-decomposed mode shows structures qualitatively similar in scale to those of the present method.
In contrast, the scale-decomposed field for $\sigma_1=\sigma_{\rm max}/6$ contains the rib structures in addition to the dominant vortex cores, suggesting that the scale-based decomposition inherently conflates rib structures with vortex cores due to their overlapping length scales.
Consequently, this comparison suggests that the present method successfully distinguishes large-scale vortical motions based on {contribution to the future lift}, not solely on the length-scale information.

\begin{figure}
    \centering
    \includegraphics[width=1\textwidth]{./Fig10.pdf}
    \vspace{-4mm}
    \caption{
    {
    $Q$-$R$ distributions of three-dimensional separated turbulent wake colored by ($a$) input and ($b$) informative $Q$-criterion.}
    }
    \vspace{-2mm}
    \label{figQ_R}
\end{figure}
\begin{figure}
    \centering
    \includegraphics[width=1\textwidth]{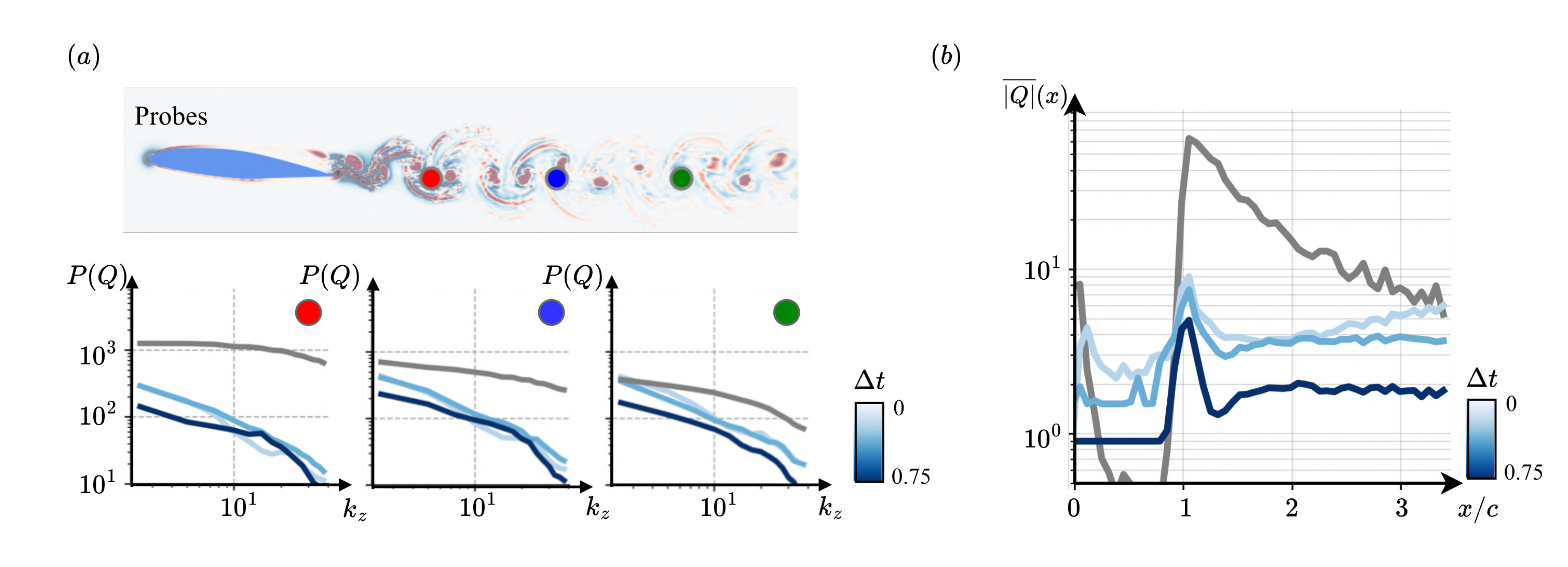}
    \vspace{-4mm}
    \caption{{Assessment of informative mode on the time window $\Delta t$, ($a$) power spectra density, and ($b$) streamwise variation of the magnitude of $Q$-criterion field {(gray: input)}.}}
    \vspace{-3mm}
    \label{fig_3Ddelt}
\end{figure}

{
Moreover, we evaluate the decomposed structures on the $Q$-$R$ plane, {where $R$ is the third invariant of the velocity gradient tensor,} as exhibited in figure~\ref{figQ_R}. 
While the second invariant $Q$ characterizes the balance between the rotation and strain rates, the third invariant $R$ governs the dynamics of vortex stretching and compression. According to these invariants, the local flow topologies are classified into four distinct states: vortex stretching ($Q>0, R>0$), vortex compression ($Q>0, R<0$), axial strain ($Q<0, R>0$), and biaxial strain ($Q<0, R<0$)~\cite{davidson2015turbulence}.
While the input field exhibits the well-known teardrop shape spanning both positive and negative $Q$-criterion~\cite{ooi1999study,fukami2024data_pi}, the present extraction is dominantly activated for rotation-dominated structures, in particular, those experiencing vortex compression.
Note that since the present model takes only the $Q$-criterion field as input, the present model selectively distinguishes between vortex stretching and compression based on the flow of information with respect to future lift, indicating a relationship with the lift-generating mechanism in the current flow field.
}

{
We then discuss the dependence of informative mode on the time window $\Delta t$.
The spanwise power spectra density of the $Q$-criterion, averaged over time, is examined in figure~\ref{fig_3Ddelt}($a$).
The present model generally underestimates informative components compared to the reference field, suggesting that redundant components are removed due to their smaller contribution to the lift generation.
Furthermore, the influence of the time delay $\Delta t$ on the power spectral density of the informative mode is predominantly concentrated in the low-wavenumber region. This implies that the contribution to the future lift is primarily governed by these large-scale structures.
}

To discuss the convection process of the vortex structure captured in extracted modal structures, the streamwise variation of the absolute value of $Q$-criterion, $\bar{|Q|}(x) = \overline{|Q(x,y,z,t)|}^{y,z,t}$, measuring the ensemble average over time and the $y-z$ plane, is assessed, as shown in figure~\ref{fig_3Ddelt}($b$).
The indicator of $\bar{|Q|}(x)$ downstream region decreases as $\Delta t$ increases, suggesting that the contribution of the structures there is recognized as less significant due to the convection process.
These findings support the applicability of the present method to turbulent flows, highlighting its potential to advance the physical interpretation of unsteady aerodynamics via {data}-driven modal analysis.

\section{Concluding remarks}
\label{sec:conc}
This study {considered} an information-theoretic machine learning method that provides time-varying informative vortical structure {related to} the future lift coefficient. 
Through three example flows: extreme vortex gust-airfoil interactions, experimentally-measured transverse gust encounter, and separated turbulent wake, the present method extracts {structures associated with lift generation from the snapshot data.}
{Built upon the previous studies \cite{arranz2024informative,fukami2025information}, which perform the decomposition locally, the present data-driven approach enables the extraction of the dynamic relationship between vortical structures and the lift response by framing the cause-and-effect association within the lift-generation mechanism.}
The current convolutional network-based deep sigmoidal flow decomposes the input flow field globally with a spatiotemporal arrangement of vortical structures preserved.
Furthermore, a low-order representation of the informative vortical structures and the aerodynamic coefficient is identified while performing decomposition.
The proposed method is capable of visualizing the relationships and offers key insights into the mechanism of force production under highly unsteady aerodynamics, based solely on flow field data and information metrics.

{
The current global extraction approach enables the extraction of coherent structure across the snapshots since the present information-theoretic convolutional machine-learning model has access to the spatial arrangement of vortical structures. 
Due to the capability of the present method to account for such spatial correlations, one promising future prospect is to apply this global approach to flow fields characterized by a broader range of structural scales, including wall-bounded turbulence. Furthermore, it will be of interest to conduct a direct comparison between the present global framework and the conventional local decomposition \cite{arranz2024informative,fukami2025information}. By examining how the incorporation of spatial arrangements influences the mode extraction process compared to an isolated point-wise analysis, we could further discuss the coupling effects of vortical structures on the target variable.
}

{We interpreted the lift-generation mechanism as a cause-and-effect relationship, where vortical structures act as the cause, while the resulting aerodynamic force as the effect.
To elucidate this relationship from an aerodynamic approach, one can consider using the force elemental method~\cite{chang1992potential} or the vortex force map approach~\cite{otomo2025vortex}. 
Comparing or incorporating them would be of interest to deepen our understanding of vortex-induced lift generation.
{Since how the input and output, along with the optimization setup, are prepared greatly affects the resulting structures assessed as informative~\cite{cremades2025additive, hoyas2025deep}, examining the dependence of identified structures on the selected variable and the mathematical definition of the cause-and-effect relationship would be of interest.
Although this study considered the relationship between lift and vortical structures over time based on prior knowledge of aerodynamics, evaluating them based on emerging approaches in the field of causal inference may clarify the rather ambiguous concepts of causality in the current vortex-induced force generation problem~\cite{imbens2015causal}.}
}

{Furthermore, the application of extracted ``informativeness" in terms of information theory can be explored by incorporating the current approach to the reinforcement learning-based active flow control. While a previous study has utilized informative components for instantaneous opposition control for drag reduction in turbulent channel flows~\cite{arranz2024informative}, extending this concept to lift-enhancement strategies will provide insights for aerodynamic flow control.
}

\section*{Acknowledgments}
{
K.F. acknowledges support from the JSPS KAKENHI Grant No.~JP25K23418, the JST PRESTO Grant No.~JPMJPR25KA, and the MEXT Coordination Funds for Promoting Aerospace Utilization Grant No.~JPJ000959. 
R.A. acknowledges support from the JSPS KAKENHI Grant No.~JP24K22942 and the JST PRESTO Grant No.~JPMJPR25K1.
K.F. and R.A. acknowledge support from the JSPS KAKENHI Grant No.~JP26K01129.
Q.L. acknowledges support from the US AFOSR Grant No.~FA9550-24-1-0069.
}

\section*{Declaration of interests}

{The authors report no conflict of interest.}


\bibliographystyle{unsrt}  
\bibliography{refs}

\end{document}